\documentclass[%
 reprint,
superscriptaddress,
 amsmath,amssymb,
 aps
]{revtex4-2}

\usepackage{graphicx}% Include figure files
\usepackage{xcolor}
\usepackage{physics}
\usepackage{amsfonts,mathtools}
\usepackage{amsthm}
\usepackage{CJKutf8}
\usepackage{physics,dsfont}
\usepackage{lipsum}
\usepackage{multirow}
\usepackage{graphicx}% Include figure files
\usepackage{dcolumn}% Align table columns on decimal point
\usepackage{bm}% bold math
\usepackage[colorlinks=true,linkcolor=blue,citecolor=blue]{hyperref}% add hypertext capabilities
\usepackage{algorithm2e}

\usepackage{caption}
\usepackage{subcaption}
\usepackage{enumitem}

\definecolor{bdf}{RGB}{49, 140, 231}

\newcommand{\revise}[1]{{\color{black} #1}}
\newcommand{\revisefinal}[1]{{\color{black} #1}}

\hypersetup{
    colorlinks=true,
    linkcolor=blue,
    filecolor=magenta,      
    urlcolor=blue,
    pdftitle={Overleaf Example},
    pdfpagemode=FullScreen,
    }

\begin{document}

\title{Emergence of fluctuation relations in UNO}% Force line breaks with \\

\author{Peter Sidajaya}%
\email{peter.sidajaya@u.nus.edu}
\affiliation{Centre for Quantum Technologies, National University of Singapore, 3 Science Drive 2, Singapore 117543, Singapore}

\author{Jovan Hsuen Khai Low}%
\affiliation{%
 Eunoia Junior College, 2 Sin Ming Place, Singapore 573838, Singapore
}%

\author{Clive Cenxin Aw}%
\email{ccx.aw@nus.edu.sg}
\affiliation{Centre for Quantum Technologies, National University of Singapore, 3 Science Drive 2, Singapore 117543, Singapore}

\author{Valerio Scarani}%
\affiliation{Centre for Quantum Technologies, National University of Singapore, 3 Science Drive 2, Singapore 117543}
\affiliation{Department of Physics, National University of Singapore, 2 Science Drive 3, Singapore 117542}

\date{\today}% It is always \today, today,

\begin{abstract} 
\revise{In the last two decades, fluctuation theorems have been proved formally and demonstrated experimentally for several variables (such as entropy production, work, or flux) and different noises causing the fluctuations (of either thermal or other origin; Markovian or non-Markovian). Here we report the observation of a detailed fluctuation relation in a statistical process outside thermodynamics and physics: the card game UNO. As the fluctuating variable, we consider the number of steps $W$ needed for one player's deck to change from $x$ to $y$ number of cards. The other players and the remaining cards play the role of a finite non-Markovian bath. Numerical simulations of runs of the game show that $W$ obeys a fluctuation relation analogous to Crooks' theorem. While the observed behavior shares some common features with infinite random walks, it also exhibits deviations that are clear signatures of non-Markovianity and the finiteness of the bath: Notably, the parameter corresponding to temperature depends strongly on the transition $x\rightarrow y$. Our paper contributes to extending the scope of fluctuation theorems beyond their usual thermodynamic setting.

%While applicable in various physical contexts, fluctuation relations are well-known as generalisations of the second law, describing relationships between work, temperature, and free energy in thermodynamic processes. They have been used extensively in studies of irreversibility and entropy, and many experiments have verified these relations for a variety of systems. In this study, we observe the same behavior beyond thermodynamics and physical contexts. In particular, we find non-trivial instantiations of behavior characteristic to Crooks fluctuation theorem for the card game UNO. For the fluctuating parameter (usually labelled as work), we consider the number of steps one player needs to effect some choice of transition in her deck, while the other players and the remaining cards play the role of a finite, non-Markovian bath. These observations lead to insights pertaining to the emergence of Crooks-like behaviors for random walks as well and the role of non-Markovianity in such contexts.
}
\end{abstract}

\maketitle

%\tableofcontents

\section{Introduction}

Fluctuation relations have become a mainstay in a variety of fields in science. \revise{Initially derived for linear response, these relations were later extended to nonlinear systems \cite{bochkov-kuzovlev-1977} and are now a key part of stochastic thermodynamics \cite{crooks-reversal,crooks-theorem,jarzynski-review-2011, VDB2013stochastic,evans2002fluctuation,harris2007fluctuation}, featuring also in fields like biophysics, optomechanics, and quantum thermodynamics \cite{exp-Collin05,exp-hayashi2018application,exp-hoang2018experimental,exp-kiesel22, tasaki2000jarzynski,campisi-haenggi-review-2011,aberg-quantum-fluct}.} 
Their significance lies in their generalization of the second law, being applicable even when systems are driven far out-of-equilibrium under nonquasistatic transformations. They describe how, despite a lack of thermodynamic constraints, strong relationships connecting average work, temperature and free energy hold. For this reason, fluctuation relations are invoked as measures of irreversibility and entropy, beyond the thermodynamic regime \cite{campisi-haenggi-review-2011,Seifert_2012,gawedzki2013fluctuation,Funo2018}.

It has also become widely accepted that much of the underlying structure of fluctuation relations comes from logical and mathematical necessity \cite{BS21,AwBS, hack2022logic}. The physical features of the system and whether some scenario adheres to some specific rendition of a fluctuation relation (say, Crooks' theorem) boil down to assumptions. These are expressed in terms of detailed balance, or, quite equivalently, the encoding of a reference prior that connects the forward and reverse processes \cite{VDB2013stochastic,AwBS,hack2022logic,spinney2012peda}. 
%Fluctuation relations like Jarzynski's equality hold because of how the reverse and forward processes relate to each other, alongside the fact that these distributions are normalized. 

%All this to say that once some physical scenario takes the appropriate assumptions, a Crooks graph will necessarily be plotted. When the Crooks relation cannot be plotted, then it is deduced that the physical scenario does not adhere to the assumptions that have been expected. In such situations, additional terms are sometimes introduced to correct the disparity. \clive{check Hatano-Sasa is appropriate here, if not just drop it.}

Now, with this in mind, the following question emerges: If it is the case that fluctuation relations come from logical consistency, with the physics embedded in priors and assumptions—then it seems reasonable to expect some \textit{non-physics scenarios to obey certain fluctuation relations}. There are plenty of possible contexts that may answer this question. For this paper, we focus on \textit{games}. In particular, an extensive-form scenario with moves-by-nature, expressed as a card-shedding game \cite{GTfudenberg1991game, GTwatson2002strategy}. We are talking about the popular card game, UNO. We plot fluctuation relation graphs for various transitions in hand sizes, in the context of this game. In particular, we found that when a specific class of cards are omitted from the game, most transitions adhere to Crooks' fluctuation theorem. 

We section the paper in the following way:
\begin{enumerate}[label=(\roman*)]
    \item In Sec. \ref{sec-uno}, we review the rules for UNO. For those who are familiar with how the game works can skip this section entirely.
    \item In Sec. \ref{sec-method}, we do two things. First, we run through the conceptual building blocks of a fluctuation relation experiment. Secondly, we show how we translate this to the game-theoretic context of UNO. We also clarify how our results are obtained in a nontrivial way.
    \item In Sec. \ref{sec-results}, we include the key results to be discussed later, noting standout features that should be explained.
    \item In Sec. \ref{sec-discuss}, we provide explanations for the features raised in the previous section. In particular, we recontextualize thermodynamic notions (like temperature) to the game-theoretic setting.
    \item Finally, in Sec. \ref{sec-concl}, we consolidate our key takeaways and conclude. 
\end{enumerate}

% So far in the literature, fluctuation theorems have been discussed mostly in the context of statistical thermodynamics. This makes sense, of course, because the concept of work, free energy, and temperature is encoded directly into the relation itself. However, one might wonder whether the relation is essentially thermodynamic in nature or whether it can be found in a different setting outside of thermodynamic. To find out more, we move our attention into a class which would not be considered thermodynamic by most: games.

% When looking for fluctuation relations in different games, an important requirement to look for is the ability for the game to revert back to a previous state. For example, it is impossible to do so in chess as the number of pieces on the board strictly goes down in time. Without the ability to revert back, it is useless to talk about a reverse process as they are strictly impossible by the rules of the game. Most board games such as chess, reversi, and backgammon fall into this category of non-reversible games. Fortunately for us, a class of card games called card-shedding games, however, offer the possibility of going back in circles. These are games whose objective is to lose all of one's cards, while randomness and the other players would sometimes force the player to draw more cards. In theory, these games could run forever, if the players were unlucky enough. Out of this class, UNO is probably the most famous one.

\section{Reviewing the Rules of UNO}\label{sec-uno}

\begin{figure}
    \centering
    \includegraphics[width=0.5\textwidth]{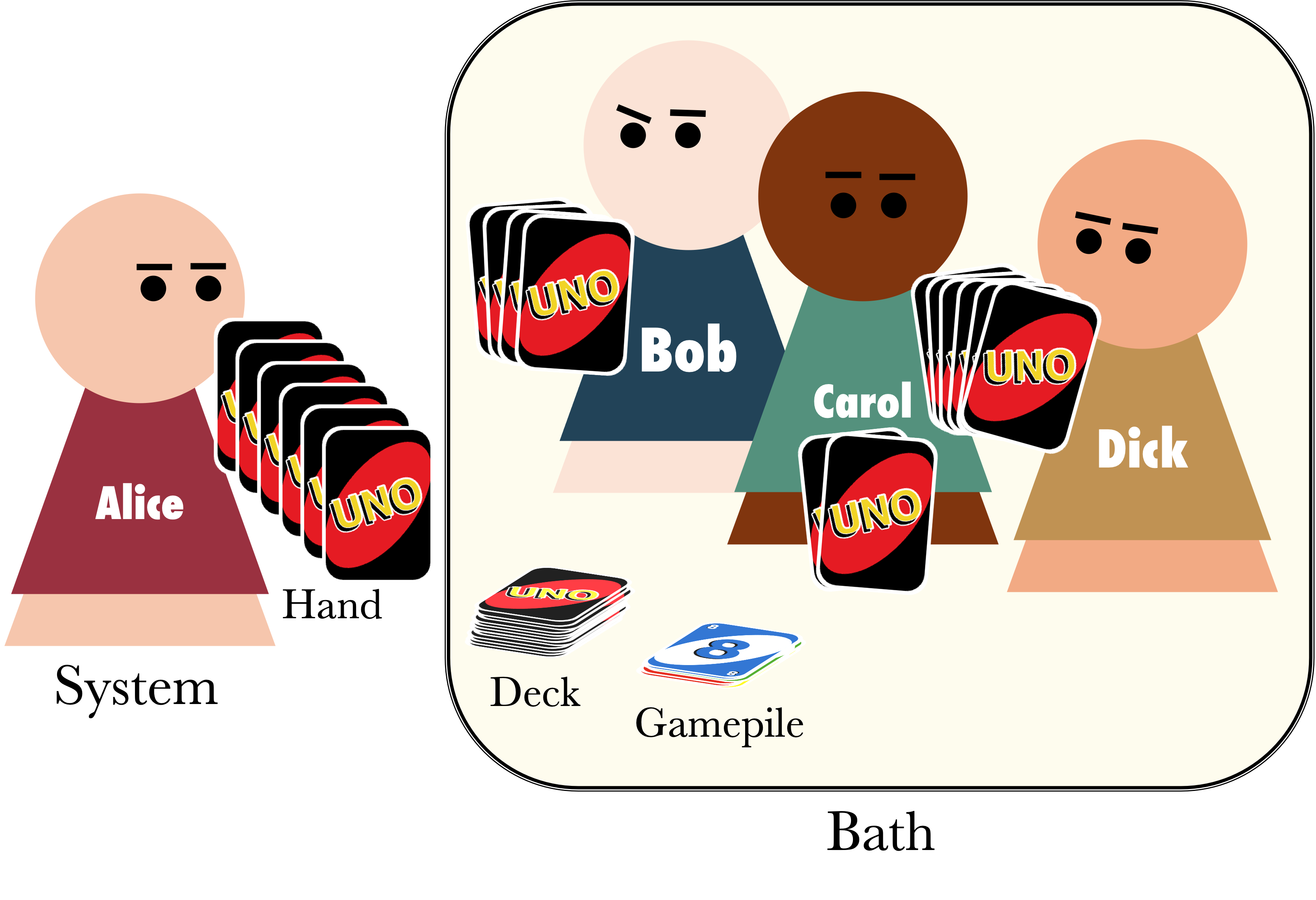}
    \caption{The setup of a typical UNO game.}
    \label{fig:setup}
\end{figure}

UNO (from Spanish and Italian for ``one''), is a proprietary American card game developed in 1971 by Merle Robbins and is currently owned by Mattel, Inc. It is a shedding-type card game where players take turns to play a card onto a game pile, with the objective of losing all the cards in one's hand. It is a derivative of a game from the 1940s known as \textit{Eights} or \textit{Crazy Eights} \footnote{These games are likewise a derivative of an even older card-shedding game called \textit{Mau-Mau}.}. These two earlier games could be played with a standard poker deck (sometimes with certain cards removed). UNO is played with a specially printed deck. There are many variations to the game and house rules are very common. In this paper, we will go with the official rules \cite{uno-instructions}.

The game consists of the deck (face-down), the gamepile (face-up), and the players' hands (known only to the respective owner). During a player's turn, they must choose and play a card from their hand onto the game pile, according to the card on top of the game pile. The card played must match either the same color, or the same number/symbol as this topmost card. If this is not possible and there are no other legal moves, the player must draw one (and only one) card from the deck. 

The game starts with every player being dealt a hand of seven cards, drawn randomly from the deck. The topmost card from the remaining deck is drawn onto the game pile. Once the starting player is designated, the turn order begins clockwise \footnote{There are official rules about how the dealer and the first player is chosen and so on. There are also rules about what to do when your handsize is reduced to one. These are irrelevant for this paper.}. The first player to lose all their cards wins. 

We move on to the deck composition. The standard UNO deck consists of 108 cards:
\begin{enumerate}[label=(\roman*)]
    \item \textbf{Normal cards}: 
    These cards are separated into four colors (blue, green, red, and yellow). Each card has a number between 0 to 9 and there are two copies of each, except for 0, which only has one, for a total of seventy-six cards.
    \item \textbf{Special cards}:
    These cards come in the same four colors but they enforce special actions or rulings when played. These come in three types: \begin{enumerate}[label=(\alph*)]
        \item The skip card skips the next player.
        \item The reverse card reverses the direction of the play, i.e., it inverts the turn order.
        \item The +2 card forces the next player to draw two cards from the deck and skip their turn.
    \end{enumerate} There are two copies of each, for a total of twenty-four cards.
    \item \textbf{Wildcards}: These cards do not feature any specific color or number. Rather, they allow the player to decide the next color to be matched. The first variation of wildcard does just this, while a second variation also invokes the ruling of the +2 card except that it adds four cards instead of two. During a player's turn, they can always play a wildcard in their hand regardless of what the game pile's topmost card is. There are four copies of each variation, for a total of eight cards. 
\end{enumerate}
Of special note are the +2 cards and the +4 wildcards. We shall refer to them as \textbf{plus-cards} in this paper. They will be shown to affect how much the game adheres to Crooks' fluctuation relation, which we will now introduce.

% For the rest of the main text of the paper, we will discard the +2 action cards and the +4 wildcards because these plus-cards add a layer of complexity that brings the game away from the basic card-shedding form of the game without adding anything interesting in terms of the fluctuation observed. However, we still run simulations of UNO games with these cards and a further discussion of these cards and their effects can be found in Appendix~\ref{app:plus}.

\section{Methods}\label{sec-method}
\subsection{Key elements of a fluctuation relations experiment}

\revise{The first \textit{detailed fluctuation theorem} was derived for the thermodynamic work \cite{bochkov-kuzovlev-1977}. Later, similar expressions have been derived for other fluctuating variables: for instance, entropy production \cite{evans2002fluctuation,Esposito2010} or flux in a diffusion process \cite{seitaridou2007measuring}. Our paper being about a game of cards, it is agnostic as to the physical analogy that we adopt. For definiteness, we refer to possibly the most famous detailed fluctuation theorem, derived for fluctuating work by Crooks in 1998 \cite{crooks-theorem}:}
\begin{equation}\label{eq:CFT-crooks}
    \frac{P_f (W)}{P_r (-W)} = e^{\beta (W-\Delta F)}.
\end{equation}
On the left-hand side is the ratio between the distribution of work $W$ observed in the \textit{forward} process and its associated \textit{reverse} process, while the right-hand side relates to the physics of the process. Crooks derived it for a purely deterministic (Hamiltonian) drive between two Gibbs state with inverse temperature $\beta$ and difference in free energy $\Delta F$. The same expression was later derived for the isothermal drive (the system undergoing evolution generated by a Hamiltonian drive while exchanging energy with a thermal bath). This became the standard introductory scenario, see e.g.,~\cite{VDB2013stochastic}. Related detailed fluctuation theorems appear in many varieties: for instance, by conditioning on an observed value of some other parameter \cite{jarz2000}. The version of Crooks' theorem that we shall use is \begin{equation}\label{eq:CFT-prop}
    \frac{P_f (W|x\rightarrow y)}{P_r (-W|y\rightarrow x)} = e^{\beta (W-W_0(x,y))},
\end{equation}
where $x$ is the initial (final) state of the target system for the forward (reverse) process, $y$ is the final (initial) state; $W$ designates work done \textit{on} the target; $W_0(x,y)$ is an offset for the origin of work ($\Delta F$ in thermal setting, but we would not have this interpretation here). 

% valerio{Should we already write here $\beta(x\to y)$, and anticipate that we won't have a single ``temperature''? Also $W_0$ may depend on $x,y$ by the way.} Reply: I think introducing the fact that our `temperature' is not really a temperature and changes with the transition at this point is too early and will confuse the readers.

% [I think that we are plotting the conditionals $P_f (W|x,y)$ rather the total $P_f (W,x,y)$, as written in the paper. I know it's eventually the same number...]

In fact, any variable (typically entropy production), that can be written as the log-ratio between the statistics of the forward and the reverse processes, automatically obeys the detailed fluctuation theorem \cite{Seifert-2005,AwBS,Auffeves_2021}. To avoid falling into this triviality, experimental tests of detailed fluctuation theorem must define independently the variable to be measured (``work'') and the recipe for the reverse process. In the paradigmatic biophysical setting \cite{exp-Collin05}, the forward process is pulling a folded RNA strand until it straightens, the reverse process consists in pushing it; meanwhile, the work made on the system is measured as the energy invested to make the transition happen. We shall follow similar prescriptions in this paper.

\subsection{Getting fluctuation relations for UNO}

\begin{figure}
    \centering
    \includegraphics[width=0.5\textwidth]{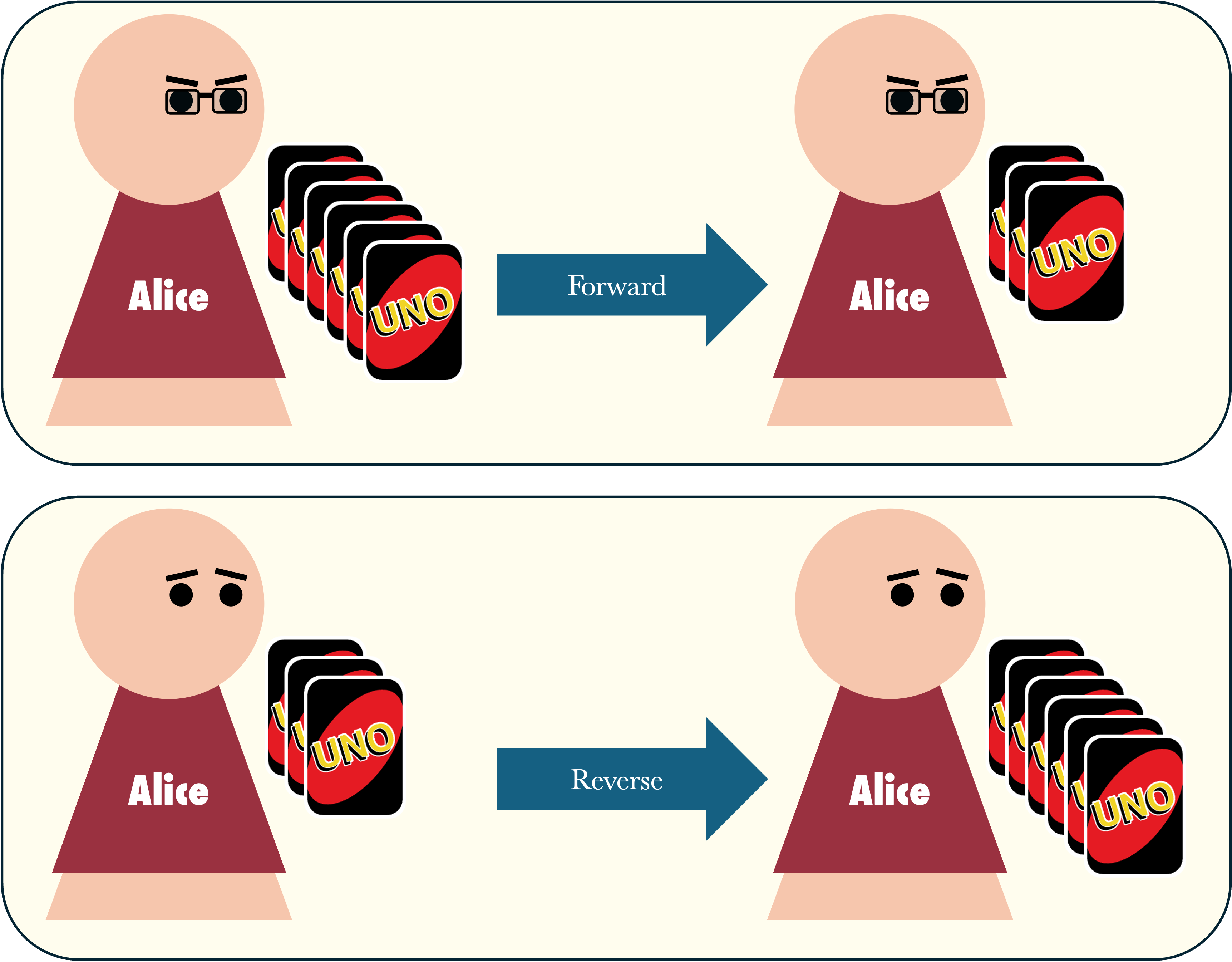}
    \caption{In the forward process, Alice follows a protocol that is effective at shedding cards. Meanwhile, in the reverse process Alice does the opposite of whatever the forward protocol entails. See Sec.~\ref{ssec-forwardreverseprocess} and Appendix \ref{app:strat} for details.}
    \label{fig:setup}
\end{figure}

\begin{figure*}[]
    \centering
    \begin{subfigure}[]{0.5\textwidth}
        \centering
        \includegraphics[width=\linewidth]{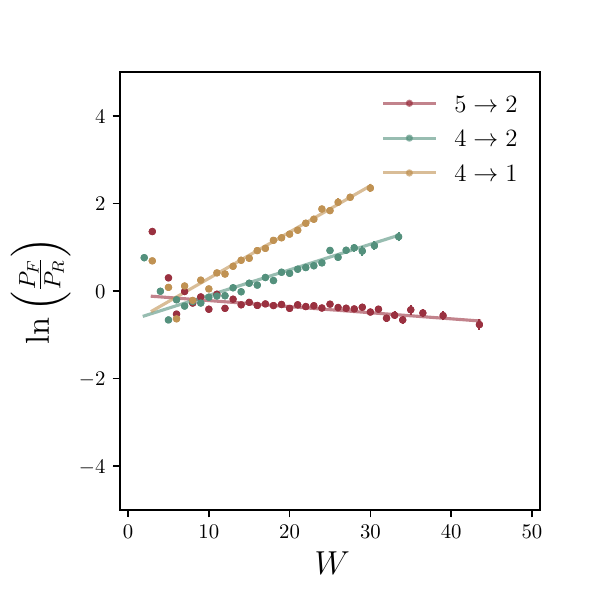}
        \caption{}
        \label{fig:Crooks plus} 
    \end{subfigure}%
    ~ 
    \begin{subfigure}[]{0.5\textwidth}
        \centering
        \includegraphics[width=\linewidth]{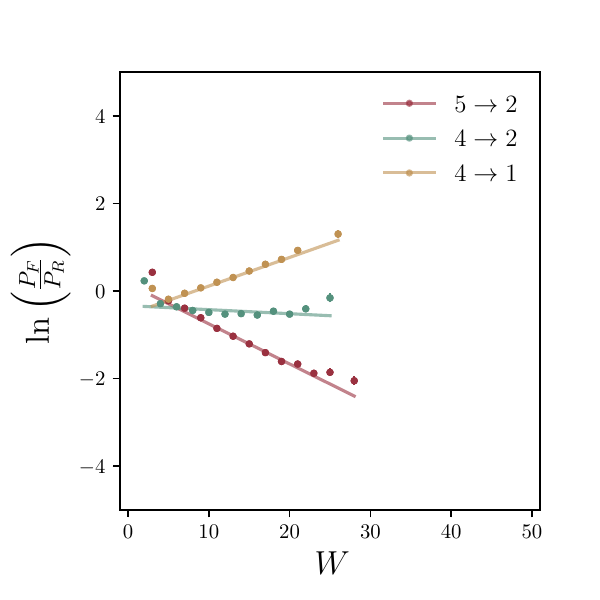}
        \caption{}
        \label{fig:Crooks no-plus}
    \end{subfigure}
    \caption{ Two sets of Crooks’ graphs for (a) UNO under official rules and (b) UNO without the +2 and +4 cards. The data used to plot the graphs consist of \revise{$10 \times 10^6$} UNO games, split evenly between the two different versions of the game and the forward and reverse processes in each of them. The lines are the Theil–Sen regression lines for each of the transitions, which is more robust to outliers than the simple linear regression \cite{theil1950rank, sen1968estimates}. \revise{Variable binning is used for values with less than 100 occurrences \cite{adami2002critical}. The deviation from the straight line at small values of $W$ is relevant; that at large values is likely caused by insufficient sampling.}}
    \label{fig:Crooks}
\end{figure*}

Having reviewed the key elements of fluctuation theorem experiments, we move to translate them into our UNO setting.

\subsubsection{Target system and bath}
For studies of fluctuation theorems, the target system and the heat bath need to be identified. This distinguishes the forward and reverse processes, demarcating what deterministic work is being done on and where stochastic heat is coming from. For the experiment in \cite{exp-Collin05}, the target and the bath would be the RNA strand and the wet environment respectively.

For this study's UNO games, we can isolate one player, labeled as Alice, as the target system. The game pile, main deck, and the other players form a stochastic context that acts as a bath contributing randomness into Alice's states. It is a finite bath with memory, i.e.~non-Markovian, because the deck is finite and (anti)correlated with the system, that is, the cards comprising the bath are those that are not with Alice.

In our simulations, all the other players in the bath are playing randomly, as opposed to Alice, which always play with a strategy (see Sec.~\ref{ssec-forwardreverseprocess}). The impact of changing the behavior of the other players can be seen in Appendix~\ref{app:smart bath}.

\subsubsection{Initial and final states}
What are ``Alice's states''? The most straightforward option is to use the set of the cards in Alice's hand as the states. However, such a definition would lead to a state space that is too vast to work with as the number of possible combination of cards is extremely large. Thus, we choose the handsizes (that is, how many cards are in Alice's hand) as the states $x,y$ that parametrises the forward and reverse statistics. This designation can be seen as a coarse graining of the exact microstates of the system (that being the exact set of cards at hand), which is standard when defining states in physical experiments \cite{exp-Collin05, exp-carberry2004fluctuations, exp-hoang2018experimental}.

%\clive{@ Peter, if you want to mention coarse-graining, it makes sense to compare this to other FT experiments, such as the Collins experiment of RNA. It's no as if unfolded and folded are the exact microstates. They are a coarse-graining also.}

\subsubsection{Fluctuating variable (``work'')}
\revise{As fluctuating variable $W$} we take the number of deck interactions Alice makes to go from $x$ to $y$ ($y$ to $x$) for the forward (reverse) process. That is, whenever she needs to draw cards, $W$ increases by 1. Whenever she places a card, $W$ increases by 1 as well. \revise{This fluctuating variable is very similar to a first passage time: we explore the connection with this well-studied notion in a Markov chain in Sec.~\ref{sec-discuss:random}. As mentioned, we shall call this fluctuating variable ``work'', motivated ultimately by a narrative connection with Crooks' theorems. That being said, in experiments, the work performed on the system is directly related to the time taken to effect the desired transition; and this connection did inspire our choice of the definition of the fluctuating variable. Also, in a thermal setting, work fluctuates because of the stochastic environment. Likewise, in our card-shedding context, Alice's deck interactions fluctuate based on her decisions and the randomness coming from the other players and the deck. By definition, our fluctuating variable cannot take negative values. Thus, to maintain Crooks' notation and compare $P_F(W)$ with $P_R(-W)$ instead of $P_R(W)$, we simply count deck interactions in the reverse process with negative integers.}

\subsubsection{Forward and reverse process}\label{ssec-forwardreverseprocess}

Finally, we need to define the forward and reverse processes. The forward and reverse probabilities are differentiated not only by their trajectories (i.e., $x \longrightarrow y$ versus $y \longrightarrow x$), but also the reversal of the work protocol done in both processes. \revise{Note that the work protocol, sometimes also referred to as the process variable, is distinct from the \textit{work} discussed in the previous section. The work protocol represents the action of the experimenter on the system and is thus deterministic and under the direct control of the experimenter. The work variable is what is measure, the stochastic fluctuating variable.}

\revise{Therefore}, in a similar vein to other experiments with fluctuation relations, we characterize forward and reverse processes by the agent's (Alice's) actions on the system. In our context, what defines the forward or reverse process has to do with how Alice makes decisions, changing the state of her hand. Hence, the forward process is designated as Alice play according to a set of protocols that, statistically speaking, constitute a decent strategy at winning. For example, if Alice has a wildcard on hand, she will not play it unless she has no other legal moves. This helps her avoid the situation where she is forced to draw cards.

We then designate as reverse process the protocol that does the converse of the forward protocol. Under the reverse process, Alice plays any wildcard she has on hand as soon as she can. This gives a strategy more likely to lose (see Fig.\ref{fig:setup}). Note that under the official rules, always drawing a card every round is a valid strategy, but for less trivial behavior we shall force Alice to always put down a card whenever possible. As with the physical experiments, the work protocol (coming from the agent or experimentalist) is reversed while being done in the same stochastic environment (in this case being the other players). Details of the protocol are given in Appendix~\ref{app:strat}.

\subsection{Experimental Method}

% Before we can find fluctuation theorems in an UNO game, we first need to define the objects of interest in the game. To imitate an actual experiment, we need to define them analogously to a thermodynamic experiment. This means that we have to have the system, bath, state, and work. However, since UNO is inherently different from a thermodynamic system, some of the definitions are bound to be arbitrary. Nevertheless, such definitions still need to be made to have a fluctuation theorem.
% First, we need to define the system and the bath. For this, we take one of the players, Alice, as the \textit{system} of interest, while the rest of the player and the deck then become the \textit{bath}. Next, we define the \textit{state} as the number of cards in Alice's hand. Transitions between states are then simply transitions between different hand sizes of Alice, which are effected by playing the game. This definition of the state is actually a coarse-graining of the `actual' states of the game, which are too numerous to collect the statistics with. Finally, in an actual experiment, \textit{work} is the amount of energy put into or given by the system to effect a transition, which is a random variable due to the presence of the heat bath. In an UNO game, a close analogy of this would be the number of turns that Alice has to make to make a certain transition. This parameter is also a random variable due to the rules and the actions of the bath.

We perform our experiment by simulating \revise{$N=10 \times 10^6$} UNO games on a computer and analyzing the data of the results. Half of the games are played with normal UNO rules while in the other half the plus-cards (the +2 cards and +4 variation wildcards) are removed. Moreover, in half of the games Alice follows the forward algorithm, in the other half the reverse algorithm. In analyzing a certain transition $x \rightarrow y$ we first obtain the probability distribution by sampling from the trajectories of Alice's hand size in each game in the forward process games whenever this transition occurs, and vice versa for $y \rightarrow x$ in the reverse process games. We then compare the probability distributions $P_f(W|x \rightarrow y)$ and $P_r(-W|y \rightarrow x)$ to obtain the desired plots corresponding to the log of the ratio $P_f(W| x \rightarrow y)/P_r(-W|y \rightarrow x)$ against $W$. These graphs are shown in Fig. \ref{fig:Crooks}, one for ordinary UNO games and another for UNO games with the plus-cards removed. We see that Crooks' relation is overall obeyed, with features to be analyzed next.

%Before that, it is important to explain why our observation is not a triviality, as there is growing awareness that fluctuation theorems are, in some respect, mere mathematical identities \cite{Seifert-2005,AwBS,Auffeves_2021}. In particular, the distributions of a random variable of the form $\sigma_{x,y}(W) = \ln P_f(W,x,y) / P_r (-W,y,x)$ automatically obey Crooks' theorem. Had we \textit{postulated} that the entropy production $\sigma$ must be linear in $W$, we would have fallen into triviality. Instead, we did not postulate any form of entropy production (in fact, we did not even introduce that notion): but we \textit{observed} after experiments that the relation between $W$ and $\ln P_f(W,x,y) / P_r (-W,y,x)$ is linear.}

%It could be necessary to clarify what we are \textit{not} doing. We are \textit{not} collecting statistics for $P_f(W,x,y)$, defining an entropy production $\sigma_{x,y}(W)$ that is linear in the parameter $W$, in order to simply reverse-engineer $P_r$ statistics based on the definition $\sigma_{x,y}(W) = \ln P_f(W,x,y) / P_r (-W,y,x)$. Had we done so, Crooks' theorem would hold as a mathematical necessity \cite{Seifert-2005,AwBS,Auffeves_2021}. Rather, we have collected the statistics for $P_f$ and $P_r$ independently, and then plotted $\ln P_f(W,x,y) /P_r (-W,y,x)$ for the definition of $W$ given above. In other words, we have approached these quantities as in an actual physical experiment, albeit for a card game. 

\section{Results}\label{sec-results}

% \clive{In my mind, it could help to just include the graphs here and maybe just note features to be explained later. Then explain them inside the discussion section.}

\begin{figure}
    \centering
    \includegraphics[width=\linewidth]{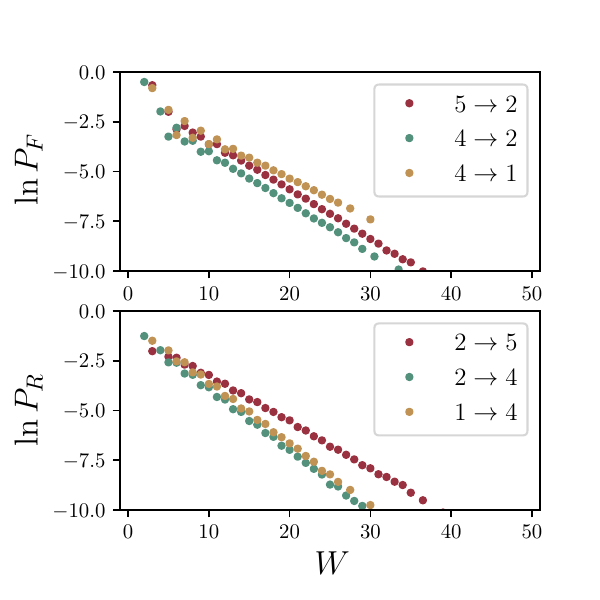}
    \caption{The probability mass functions (pmf) of different transitions for the forward \textbf{(top)} and reverse \textbf{(bottom)} processes. Similar to a random walk (see Appendix \ref{app:randomwalk}), the pmf is in the form of an exponential distribution. Here, only the transitions in the full game with the plus-cards are shown.}
    \label{fig:pmf}
\end{figure}

Figure \ref{fig:Crooks} shows the plots of $\ln{\left(\frac{P_F}{P_R}\right)}$ against $W$ from the collected statistics of the game for different transitions. There are two versions of the plot, one for the official UNO game, and another one for the same game but with the plus-cards removed. Let us first highlight some notable features observed in the figures, which we will discuss in detail in the next section. The individual probability mass functions (pmf) of the forward and reverse processes can also be seen in Fig.~\ref{fig:pmf}.

In all cases, it could be seen that a linear relation is obeyed in the $W \gtrsim 5$ regime, with a bend around $W \sim 5$, which gives a hockey-stick character to the plots. The behavior at larger values of $W$ seems to become more erratic, but this is caused by the increasingly low number of samples. Regression lines for each of the transitions are also given in the graph. Let us note here that prior to this, there is no reason to expect the linear relation found here. In fact, if we had defined the work differently (for example, by the number of wild cards or red cards used in effecting a transition), we would not see this relationship. We discuss more the underlying reason for the linear relationship in Sec.~\ref{sec-discuss:random}.

It can clearly be seen that the gradients are different for each of the transitions. Notably, the gradients can be negative. We will discuss the interpretation of this in Sec.~\ref{sec-discuss:temp}. Furthermore, there are differences between Figs.~\ref{fig:Crooks no-plus} and \ref{fig:Crooks plus}, which show the effects of removing the so-called plus-cards from the game. First, the gradients for the same $x,y$ transition decrease when such cards are removed from the game. Furthermore, for the ordinary deck, there are more data points due to the extension of the support to both even and odd $W$ values and there is a small oscillatory behavior happening in the small $W$ regime. Finally, when the plus-cards are removed, there are fewer data points in the large $W$ regime. We elaborate more on these specific points in Sec.~\ref{sec-discuss:plus}.

\section{Discussion}\label{sec-discuss}

Our numerical experiment has found Crooks' relation occurs in a card game, for a non-thermodynamical analog of work and for dissipation in a finite non-Markovian bath. First, in order to gain some intuition on why Crooks' theorem holds in such a case, we compare our process to the familiar Markovian random walk (Sec.~\ref{sec-discuss:random}). We then trace the origin of the observed ``temperature'' (be it positive or negative) to the relation between the starting point of the process and the typical hand size (Sec.~\ref{sec-discuss:temp}). Finally, we discuss how the plus-cards affect the extent in which Crooks' relation is adhered to (Sec.~\ref{sec-discuss:plus}). 

% \begin{figure*}
%     \centering
%     \includegraphics[width=0.9\linewidth]{figures/misc/markovianised.png}
%     \caption{\revise{The creation of 'Markovianised' UNO involves collecting the averaged transition probabilities from one state to another from the game data and creating a Markov chain based on those transition probabilities. In this diagram, we omit the transitions that involve plus cards, such as $P(3|1)$, for clarity.}}
%     \label{fig:markovianised-diagram}
% \end{figure*}

\begin{figure}
    \centering
    \includegraphics[width=\linewidth]{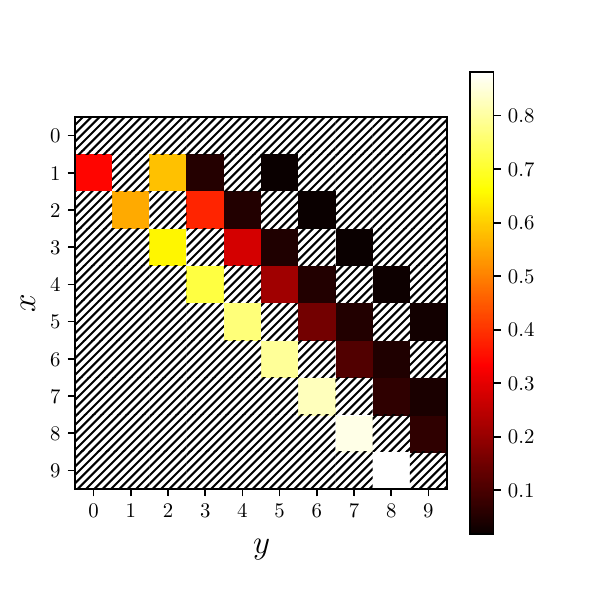}
    \caption{\revise{The colormap transition probabilities of the ``Markovianized'' UNO for the forward process games with plus-cards. The two brightest diagonal lines correspond to the transition probabilities of playing one card or drawing one card. The other two diagonals correspond to the transition probabilities of drawing multiple cards at once due to the plus-cards.}}
    \label{fig:Markov-matrix}
\end{figure}

\begin{figure}
    \centering
    \includegraphics[width=\linewidth]{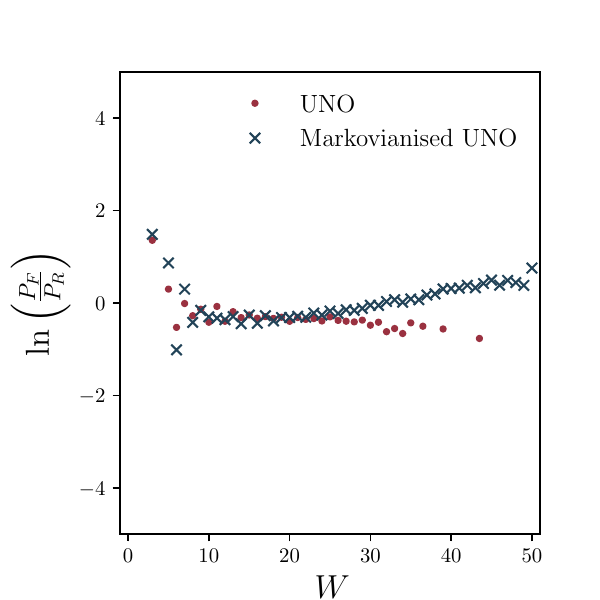}
    \caption{A comparison of $\ln{\left(\frac{P_F}{P_R}\right)}$ against $W$ before and after ``Markovianization''. The particular transition in the figure is $5 \rightarrow 2$. For other transitions, we observed qualitatively identical behaviors.}
    \label{fig:Crooks markov}
\end{figure}

\subsection{Comparison with a Markovian random walk}
\label{sec-discuss:random}

\begin{figure*}
    \centering
    \begin{subfigure}[]{0.5\textwidth}
        \centering
        \includegraphics[width=\linewidth]{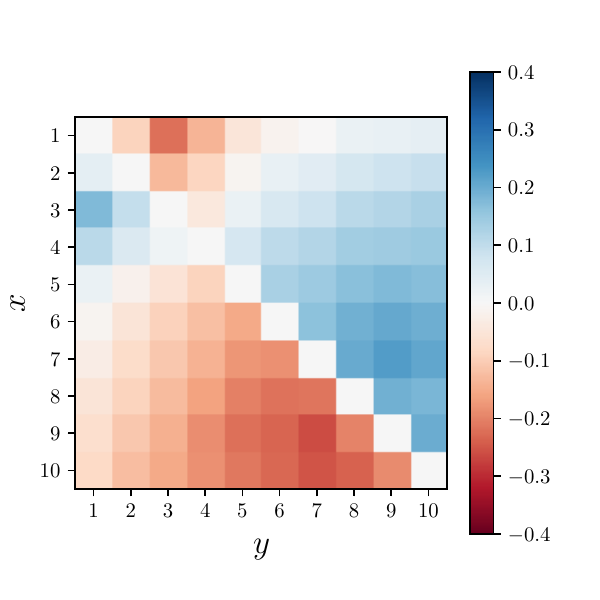}
        \caption{}
    \end{subfigure}%
    \begin{subfigure}[]{0.5\textwidth}
        \centering
        \includegraphics[width=\linewidth]{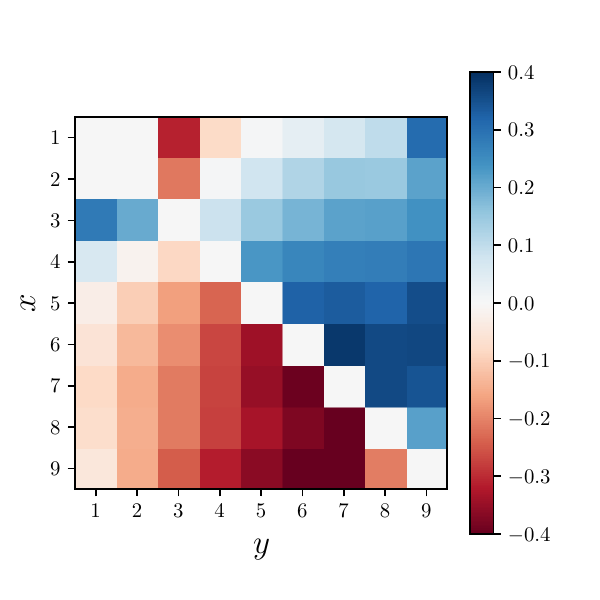}
        \caption{}
    \end{subfigure}%
    \caption{The slopes $\beta(x\rightarrow y)$ of the linear regression lines for different transitions $x\rightarrow y$, for $1\leq x,y\leq 10$ for (a) UNO under official rules and (b) UNO without the +2 and +4 cards. In the original version of Crooks' fluctuation theorem, this slope is the inverse temperature $\beta$ of the thermal bath.}
    \label{fig:regressions}
\end{figure*}

% It is clear that, given our demarcations of the system, bath and designated state space, the process of playing UNO is non-Markovian.   Indeed, the probability of losing/drawing a card on a given round depends not only on the number of cards at hand, but the actual cards themselves. Moreover, any given trajectory of an UNO game results in a memory that is embedded in the cards of the play deck, main deck and hands of other players. Transitions of the target system's handsize thus depend on this hidden memory, in a way that cannot be reduced to the previous, surface state, that it is transiting from.

{If we had defined our state space to be the full state of the game without any coarse graining, the whole process of playing UNO would be Markovian. However, because of our coarse graining process, the process is now non-Markovian. The probability of a given trajectory depends on a memory that is embedded in the cards of the play deck, main deck and hands of other players, which are not taken into account in the state space.

To make further analysis easier, it may be insightful to see whether the non-Markovian effect is negligible. Thus, we compare the original UNO process to a `Markovianised' approximation. \revise{The Markov transition matrix $M_{x,y} = P(y|x)$ is obtained from the data by averaging the transition probability between different hand sizes $x \longrightarrow y$ in a single round. In actual play this probability depends variables other than Alice's current hand size, such as the type of cards in her hand, but the 'Markovianised' approximation effectively ignores these other variables. What we obtain from this matrix is a Markov chain built from the UNO statistics where the probability of losing/drawing a card only depends on Alice's current hand size. The transition probabilities can be seen in Fig.~\ref{fig:Markov-matrix}}

The Markov chain thus obtained is a finite, nonuniform random walk process (with smaller ``jump'' transitions allowed by the plus-cards) \revisefinal{with an absorbing boundary at 0. As seen from gradient of color in the transition matrix of Fig.~\ref{fig:Markov-matrix}, the probability of playing a card in a given round is monotonically increasing with the number of cards on hand. This is intuitive as it is more likely that Alice is able to play something when she has 10 cards compared to when she only has three. Finally, it is finite since once she reaches state 0, which acts as an absorbing state, the game ends.}

\revisefinal{Our system bears resemblance to the continuous time and space Ornstein-Uhlenbeck process, which can also give an exponential first passage times under certain conditions \cite{larralde2004first}. However, fully analytical treatment of a stochastic process such as these is non-trivial and lies outside the scope of the paper. For a more complete treatment of first passage times in stochastic processes, we refer the readers to references such as \cite{redner2001guide}}

With this, we rerun the simulations as before, but for the Markov chain. A comparison of the Crooks' fluctuation graphs obtained before and after Markovianisation can be seen in Fig.~\ref{fig:Crooks markov}. The values are similar for small values of $W$, then differ for larger values. Thus, we see how the Crooks-like behavior emerges from the random-walk character in the intermediate $W$ regime.

% \valerio{I think we should change/check a couple of things here:
% \begin{itemize}
%     \item While it is obvious that Uno is similar to Markov for small $W$, even in that part a linear fit would give different slopes (positive for Markov, negative for Uno). Do you see a difference in slopes also for the same analysis of other transitions? If yes, we could say that the value of $\beta$ is an indicator of non-Markovianity.
%     \item Power law: we should not cite an equation of the Appendix that refers to random walks. I suggest that we provide (also as an appendix) the actual measured power-law distributions. In the main text, we can just mention that we observe a power law, similar to that of a random walk; and refer to the appendix.
% \end{itemize}}

The analogy with a Markovian random walk may provide some intuition on the origin of the Crooks' relation. Indeed, although a general treatment of a finite, nonuniform random walk is not trivial, it can be shown that Crooks' theorem holds in an infinite, uniform random walk, when work corresponds to first passage time (see Appendix \ref{app:randomwalk}). It is notable that this random-walk behavior actually produces exponential distributions in $W$ for each transition from $x$ to $y$ (see Fig.~\ref{fig:pmf}), as opposed to Gaussian distributions observed in physical fluctuation theorem experiments. This emphasizes that the settings that adhere to Crooks' and Jarzynski's theorems go beyond the typical thermodynamic case.

\subsection{The analog of temperature}
\label{sec-discuss:temp}

In the standard isothermal process, the slope of the Crooks line is given by the inverse temperature $\beta$. In our case, the ``bath'' is not a thermal bath, and this is most clearly manifested here: the slope of the Crooks line depends on the transition and may be negative: the values of $\beta(x\rightarrow y)$ are plotted in Fig.~\ref{fig:regressions}. For both versions of the game, we observe $\beta(x\rightarrow y)\approx -\beta(y\rightarrow x)$. For $x,y\gtrsim 4$, $\textrm{sgn}[\beta(x\rightarrow y)]=\textrm{sgn}[x-y]$: in words, the ``temperature'' is positive (negative) for the transitions that increase (decrease) Alice's cards. The opposite behavior is observed when $x,y\lesssim 4$.

In order to understand this behavior, we plot the histogram of our coarse-grained variable (Alice's hand size) throughout all the games, for both the forward and the reverse process (Fig.~\ref{fig:steady-state}). We observe that the most probable values are 3, 4 and 5. There is also a slight difference between the forward and reverse processes, the maximum being 3 for the former and 4 for the latter. 

With this, we can observe that the basic structure in Fig.~\ref{fig:regressions} is as follows: $\beta(x\rightarrow y)$ is positive for transitions that move the hand size away from the most probable number $\mu\approx 3\sim 4$, i.e.,~if $|y-\mu|>|x-\mu|$; it is negative for transitions that bring the hand size closer to it, i.e.,~if $|y-\mu|<|x-\mu|$. Also, $|\beta|$ is typically larger if the transition happens far away from $\mu$. 

\revise{An interesting point to note is that, in the case of an infinite uniform random walk (see Appendix \ref{app:randomwalk}), the slope of the fluctuation relations is the same regardless of the transition, as one might expect from a physical temperature. In turn, this shows why a single ``temperature'' is not expected in UNO: even when Markovianised (Sec.~\ref{sec-discuss:random}), the transition probabilities are not uniform but state dependent; and the walk is bounded on one side. We leave it for further work to explore whether the analogy with temperature, natural in the fluctuation theorem analysis, can be fruitful beyond it.}

%If we observe the gradients for transitions starting from $x=1$, the gradient reaches a minimum at $y=3$. The turning point in the behavior at hand size of 3 might indicate that the gradient has something to do with how much time does Alice spend in each of the hand sizes which is shown in Figure~\ref{fig:steady-state}. Since a game of UNO does not run forever \valerio{it may, may it not?}, this distribution is the closest thing to a steady state. We can see there that in the forward process games, hand size 3 is the most visited state, while hand size 4 is the most visited state for the reverse games. Overall, the cause and pattern of the changes in $\beta$ for different transitions can be intuitively understood from Eq.~(\ref{eq:temp}). Unlike an infinite random walk, the transition probabilities of in an UNO game changes depending on the current hand size, which creates the quasi-steady state mentioned earlier. Thus, it is to be expected that the effective temperature are different for different transitions.

\begin{figure}
    \centering
    \includegraphics[width=\linewidth]{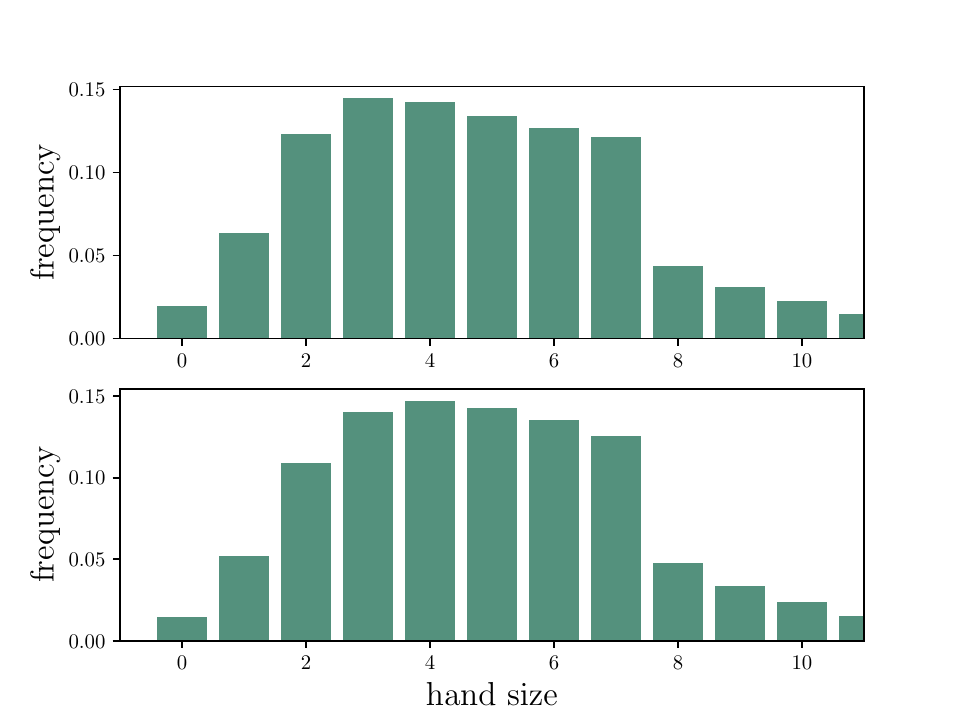}
    \caption{Histogram of Alice's hand size throughout all of the games under the complete official rules. (Top) Forward process; (bottom) reverse process. Having sampled each process for over $N=10^6$ runs, the error bars are $\sqrt{p(1-p)/N}\approx 10^{-4}$, too small to be seen at the scale of the plot.}
    \label{fig:steady-state}
\end{figure}

\subsection{The effects of plus-cards}\label{sec-discuss:plus}

As mentioned before, the removal of the plus-cards create some significant differences in the Crooks' graph, which are: the extension of the support into both odd and even numbers, the oscillatory behavior of the lines at small values of $W$, and the longer tails of the Crooks graphs.

The first two are related and can be understood by our convention to count the action of drawing due to a +2 or +4 as a single deck interaction. Because of this, transitions that used to be possible in even (odd) steps can now be achieved in odd (even) steps too—given that Alice receives a plus-card. However, the probability of receiving a plus-card is relatively small, and most transitions still happen in the parity of $y-x$. This gives an oscillatory behavior in both $P_F$ and $P_R$ between the two parity, and this behavior carries over to the Crooks graphs. This behavior, however, dies down at large values of $W$ as the chance of receiving a plus-card gets higher and higher the longer the trajectory is. This initial zigzag behavior can be seen more clearly in the pmf of $W$ in each of the processes in Fig.~\ref{fig:pmf}. Finally, the plus-cards also make it harder to shed cards regularly. Alice's hand size has a greater tendency to drift back and forth rather than decrease with regularity. This leads to generally longer paths and larger deck interactions. 

Overall, the addition of plus-cards leads the game away from a smooth behavior in the Crooks' graphs. This is expected as the plus-cards give a stepwise behavior to otherwise smooth trajectories of the hand size. 

We add that the same experiment was run for \textit{Crazy Eights} (see Appendix \ref{app-crazyeight}). Since it is played with standard playing cards without any analog to plus-cards, the Crooks plots of Fig \ref{fig:Crazy Eights} have no significant oscillatory features, as expected.

\section{Conclusions}\label{sec-concl}
% \clive{It is here that we can mention the other works that connect thermodynamics, statistical mechanics and fluctuation theorems to game theory. State that we are adding to the emerging cross-pollination between these two fields.}

We have shown that Crooks' fluctuation relation can be found in non-physics scenarios, even in a game of cards. Using the card-shedding game of UNO, we first replaced the thermodynamical notion of work with the number of interactions of one player with the deck. The bath, into which heat is gained from and dissipated into, is replaced by the cards held by the other players or sitting in the deck. Then, we defined the forward process as a behavior or protocol of a player that is relatively successful at shedding cards, and the reverse process as the converse of that forward protocol. With these definitions, we run simulations of the game and collected the statistics of ``work'' for both the forward and the reverse process. These statistics, which obey an exponential distribution show that the definition of ``work'' obeys Crooks' fluctuation theorem in these processes.

While a Crooks-like behavior is expected for a Markovian random walk, our process is non-Markovian, because the bath is finite and with memory. The unconventional nature of the bath is further demonstrated by the fact that the ``temperature'' (the inverse of the slope of the Crooks plot) can be either positive or negative, depending on the transition probabilities of the hand size. This behavior was explained as a result of the most common state being around the handsize of $3$ and $4$.

Our study establishes another bridge connecting game theory and thermodynamics \cite{ortega2013thermodynamics,adami2018thermodynamics,babajanyan2020energy,fujimoto2024game,cross-francica2024fluctuation,hack2022logic,cross-hiura2019microscopic,cross-wolpert2006information}, and provides an example of the relevance of tools from stochastic thermodynamics outside thermal physics \cite{Ducuara_23,LiuUeda23}. \revise{In particular, our study shows how such comparisons across game theory and physics produce questions that can lead to new insights---in this case, connections between random walks, their boundedness, Markovianity, Crooks' fluctuation theorem, and temperature.}

Going forward, we notice that here we fixed a simple strategy for the player and focused on establishing an experiment in analogy with thermodynamical physics. It would be interesting to see a deeper connection between different strategies of the player and their effects on the fluctuation relation. Moreover, a more extensive survey of different games, or even beyond games, would be helpful in elucidating the significance and generality of Crooks' fluctuation theorem. 

\section*{Acknowledgments} 
We thank Alexia Auffèves, Francesco Buscemi, Andr\'es Ducuara, G\'eraldine Haack, Hyukjoon Kwon, Dario Poletti, Paul Skrzypczyk and Antonios Varvitsiotis for helpful discussions and suggestions. 

\revise{This research is supported by the National Research Foundation, Singapore and A*STAR under its CQT Bridging Grant; by the same National Research Foundation, Singapore, through the National Quantum Office, hosted in A*STAR, under its Centre for Quantum Technologies Funding Initiative (S24Q2d0009)}; and by the Ministry of Education, Singapore, under the Tier 2 grant ``Bayesian approach to irreversibility'' (Grant No.~MOE-T2EP50123-0002).

This project started in the context of the Science Research Programme (SRP), a collaboration between the NUS Faculty of Science, NUS College of Design and Engineering, and the Gifted Education Branch of the Ministry of Education, Singapore. We also thank Chan Xin Yu for her support in this collaboration.

\section*{Code and Data Availability}
The codes used to run the simulations and generate the data can be found online \footnote{\href{https://github.com/PeterSidajaya/uno-fluctuation}{https://github.com/PeterSidajaya/uno-fluctuation}}. Requests and correspondences regarding the code should be addressed to P.S.

% \section*{Author Contributions}
% C.A. came up with the idea to do fluctuation theorems in games. P.S. and J.L. wrote and ran the code. V.S. provided the overall supervision to the project. P.S, C.A., and V.S. contributed to writing the paper. All authors discussed and analysed the results extensively.

% \section*{Competing Interests}
% The authors declare no competing interests.

\bibliography{apssamp}

\appendix
% \section{On Jarzynski's Equality} \label{app-jar}

% \valerio{I don't think we need this appendix... If you want to keep it, refer to it in the text again :)}

% Jarzynski's equality is easily obtained from Crooks by taking averages over the forward process:
% \begin{equation}\label{eq:JE}
%     \begin{aligned}
%         \int P_f(W) \, e^{-\beta W} \, dW &= \int P_r (-W) \, dW \; e^{-\beta \Delta F} \\
%         \langle e^{-\beta W} \rangle &= e^{-\beta \Delta F} 
%     \end{aligned}
% \end{equation}

% By ``Jensen's Inequality'' we know that the exponential of a variable's mean is never larger than the mean of the exponential of that variable:
% \begin{equation}
%     e^{\langle z \rangle} \leq \langle e^{z} \rangle.
% \end{equation}
% With \eqref{eq:JE}, this implies:
% \begin{equation}\label{eq:JE2}
%     \Delta F \leq \langle W \rangle,
% \end{equation}
% which is a statement of the second law.

\onecolumngrid

\section{Strategies used}\label{app:strat}

In order to effect the forward and reverse processes, we created two algorithms, one that aims to win (forward) and another that aims to lose (reverse). As noted earlier, it is actually legal to draw a card even when you can play a card from your hand. Thus, a possible reverse algorithm is to draw a card every round. However, to make the reverse process less trivial we will force Alice to play whenever there is a possible legal move.

One of the strategies to win UNO is to keep any wildcard on the hand for as long as possible. This is because the card can be played regardless of the state of the gamepile. Moreover, when playing a wildcard one should generally choose a color that is most numerous in one's hand. This is because color rarely changes and tends to persists between subsequent rounds. These two observations constitute the strategy that Alice uses for the forward process. For the reverse strategy, Alice tries to throw away any wildcard she has as soon as possible. Moreover, she then chooses the least numerous color on her hand, in the hope of being forced to draw a card as soon as possible.

The two algorithms are shown in Algorithm~\ref{alg:fwd} and~\ref{alg:bwd}. Against three other bots playing completely randomly the win rate of the forward and reverse processes for the official UNO game are \revise{31.7\%} and 24.7\%, respectively.

\begin{algorithm}[]\caption{The forward process.}\label{alg:fwd}
\eIf{a card can be played}
{
    \eIf{there is a wildcard on hand}
    {
        \eIf{a non-wildcard can be played}
        {
            play a random legal non-wildcard;
        }{
            $X \gets$ most common color on hand\;
            play the wildcard and declare $X$ as the next color;
        }
    }{
        play a random legal card;
    }
}{
    draw a card $Y$\;
    \If{$Y$ can be played}
    {
        play $Y$;
    }
}
\end{algorithm}

\begin{algorithm}[]\caption{The reverse process.}\label{alg:bwd}
\eIf{a card can be played}
{
    \eIf{there is a wildcard on hand}
    {
        
        $X \gets$ least common color on hand\;
        play the wildcard and declare $X$ as the next color;
    }{
        play a random legal card;
    }
}{
    draw a card $Y$\;
    \If{$Y$ can be played}
    {
        play $Y$;
    }
}
\end{algorithm}

\section{Different kinds of bath}
\label{app:smart bath}

\begin{figure*}[]
    \centering
    \begin{subfigure}[]{0.5\textwidth}
        \centering
        \includegraphics[width=\linewidth]{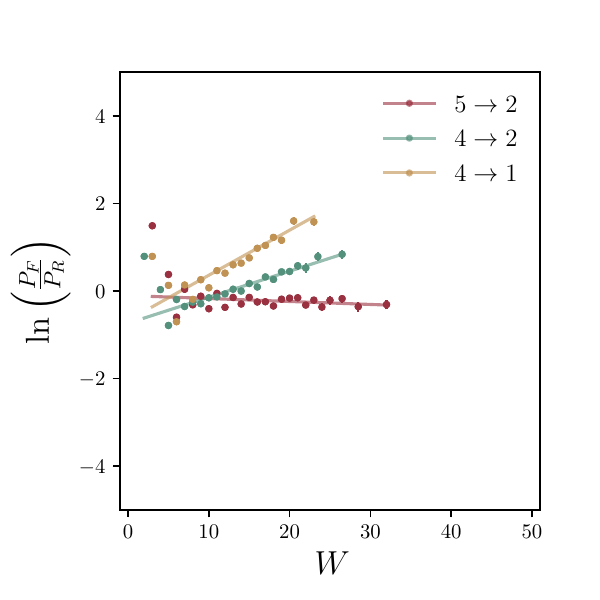}
        \caption{}
        \label{fig:smart bath}
    \end{subfigure}
    \begin{subfigure}[]{0.5\textwidth}
        \centering
        \includegraphics[width=\linewidth]{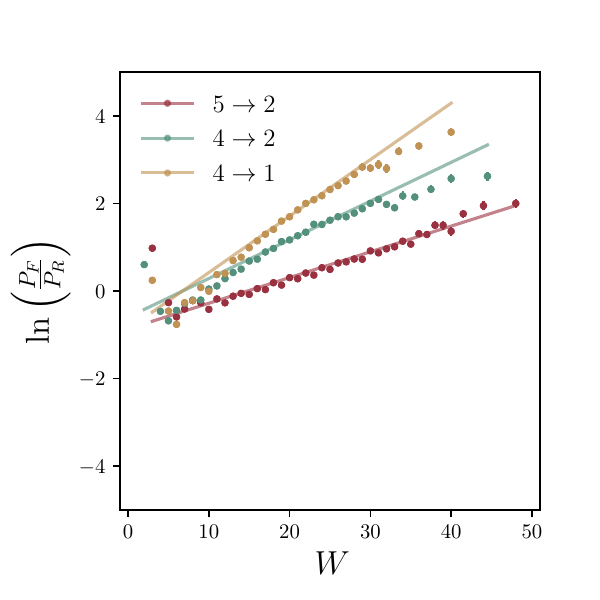}
        \caption{}\label{fig:small bath}
    \end{subfigure}%
    \begin{subfigure}[]{0.5\textwidth}
        \centering
        \includegraphics[width=\linewidth]{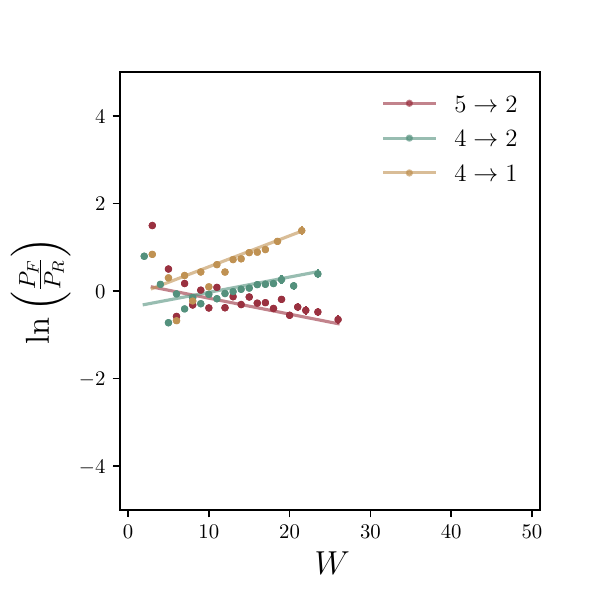}
        \caption{}\label{fig:big bath}
    \end{subfigure}%
    \caption{ The Crooks’ plot for different versions of the bath (a) UNO game with other players playing smartly (compare with Fig.~\ref{fig:Crooks plus}) (b) UNO game with two players and (c) UNO game with six players. \revise{Each of the data for these variants was obtained from two million runs.}}
\end{figure*}

In the main text, we only considered four-player games where the other players always play randomly. Here, we consider two different variations of the bath (the other players): one with smart players and another with a different number of players.

Previously, we only considered the case where the other players other than Alice always played randomly without any strategy. \revise{One might then ask the question of whether the} randomness of the other players `adds' to the stochasticity of the bath and affects our result. The short answer to this question is: No. Figure \ref{fig:smart bath} shows the Crooks' plot obtained with the same method as in the main text but with all the other players using Algorithm \ref{alg:fwd} instead of playing randomly. As can be seen, there is little to no noticeable difference with Fig.~\ref{fig:Crooks plus}.

While we only considered four-player games previously, UNO is actually designed for two to ten players, thus we also investigated what happens when we change the number of players. The Crooks' plot for a game of UNO with two and six players can be seen in Figs.~\ref{fig:big bath} and ~\ref{fig:small bath}, respectively. Two trends can be seen here: With more cards, the length of the graph shortens and the gradients for each transition go down. The difference in length can be understood from the fact that, in games with more players, the trajectories of Alice tend to be shorter. That is, in games with more players, Alice plays less in each game.

\revise{
\section{Crooks' relation for the time of first passage in an infinite random walk}
\label{app:randomwalk}

Let us consider a random walk process on the lattice of integer numbers $\mathds{Z}=\{...,-2,-1,0,1,2\}$ where transitions exist only between neighboring numbers. In each timestep, the probability of the walker going to the right ($+1$) is $p$, while the probability of it going to the left ($-1$) is $q=1-p$. In a Markov chain, the first passage/hitting time $N$ is the number of steps taken by a walker to first reach $f$ from a starting point $i$. What we are concerned with is the distribution of $P(N|i\rightarrow i+k)$, i.e., the distribution of the first passage time given that the walker has moved from $i$ to $i+k$. The closed form expression is given by \cite{grimmett2020probability}
\begin{equation}\label{eq:FPT}
    P(N=n|i\rightarrow i+k) = \frac{|k|}{n} \binom{n}{\frac{n+k}{2}}p^{\frac{n+k}{2}}q^{\frac{n+k}{2}-k},
\end{equation}
and it is zero when $\frac{n+k}{2}$ is not a nonnegative integer. This distribution is an exponential distribution (compare with in Fig.~\ref{fig:pmf}).

To make a fluctuation relation, we need two random walks $A$ and $B$ with different transition probabilities, $(p_1, q_1)$ and $(p_2,q_2)$. Taking the logarithm of the ratio of the first passage times, we get
\begin{align*}
    \ln\left(\frac{P_A(N=2n-1|i\rightarrow i+k)}{P_B(N=2n-1|i\rightarrow i+k)}\right) \\
    = \frac{n+k}{2} \ln\left(\frac{p_1q_1}{p_2q_2}\right) - k\ln\left(\frac{q_1}{q_2}\right)\,;
\end{align*}
where the analogy with Crooks with $n\equiv W$ yields
\begin{equation}\label{eq:temp}
    \beta =  \frac{n}{2}\ln\left(\frac{p_1(1-p_1)}{p_2(1-p_2)}\right), \quad \beta\Delta F = -\frac{k}{2}\ln\left(\frac{p_1(1-p_1)}{p_2(1-p_2)}\right) + k\ln\left(\frac{(1-p_1)}{(1-p_2)}\right).
\end{equation}
In particular, $\beta$ is independent of $k$, i.e., it is the same for all transitions. We have mentioned this fact in Sec. \ref{sec-discuss:temp}, and explained there why this is naturally not the case for UNO.

Finally note that, at this stage, we have not yet elaborated on the relations of the two processes. As is the case in general \cite{BS21,AwBS} (and we have explained at the end of Sec. \ref{sec-method} why our experiment does not fall into this triviality trap), the detailed fluctuation theorem follows formally from studying $\ln(P_A/P_B)$, without the need to provide any further connection between the two processes. Here, for the specific relation, we added the request that both processes be random walks.

For deeper interpretations, one may want to require further that process $B$ would be a reverse of $A$. There is no general way to reverse an abstractly defined random walk. Physically, one could think of a walk on an inclined plane (a native $p\neq 1-p$) with an additional drive, pushing in opposite directions for processes A and B (to get the effective $p_1\neq p_2$). We don't need to pursue this analogy here.
}

\section{Crazy Eights}\label{app-crazyeight}

\begin{figure*}[]
    \centering
    \begin{subfigure}[]{0.5\textwidth}
        \centering
        \includegraphics[width=\linewidth]{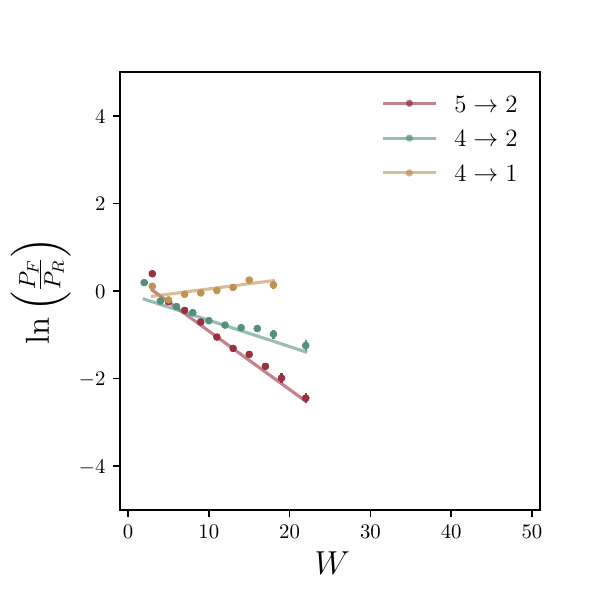}
        \caption{Crooks' plot for Crazy Eights.}
    \end{subfigure}%
    \begin{subfigure}[]{0.5\textwidth}
        \centering
        \includegraphics[width=\linewidth]{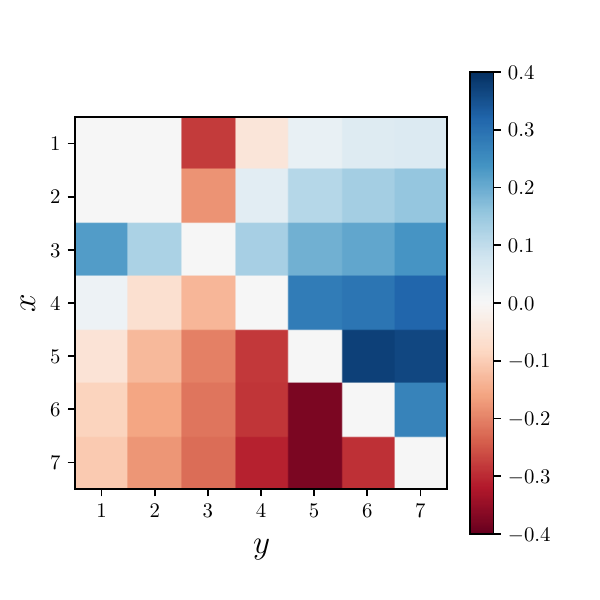}
        \caption{Regression slopes for Crazy Eights.}
    \end{subfigure}%
    \caption{The associated plots for the game Crazy Eights, on which UNO is based. \revise{Each of the data for these variants was obtained from two million runs.}}
    \label{fig:Crazy Eights}
\end{figure*}

UNO is derived from a simpler game played with the standard playing card deck called the \textit{Crazy Eights}. In the most basic version of the game, the eight cards act as the wildcards and the suites play the role of the colors. Additional homebrew rules can be added into the game that makes it closer to UNO. For completeness we create similar plots for the basic Crazy Eights game, which can be seen in Fig.~\ref{fig:Crazy Eights}. It can be seen that the plots are very similar to UNO, especially when the plus-cards are removed [see Fig.~\ref{fig:Crooks no-plus}].

\end{document}